\documentclass[aps,prd,twocolumn,superscriptaddress,10pt,noshowpacs]{revtex4}  
\usepackage[active]{srcltx}
\usepackage{graphicx}
\usepackage[utf8]{inputenc}
\usepackage{amsmath}
\usepackage{xcolor}
\usepackage{times,txfonts}
\usepackage{float}

\begin{document}

\title{Bose-Einstein condensation in Ho\v{r}ava-Lifshitz theory}

\author{J. Furtado}
\affiliation{Centro de Ci\^{e}ncias e Tecnologia, Universidade Federal do Cariri, 57072-270, Juazeiro do Norte, Cear\'{a}, Brazil}
\email{job.furtado@ufca.edu.br}

\author{J. F. Assun\c{c}\~{a}o}
\affiliation{Departamento de F\'{i}sica, Universidade Regional do Cariri, 63105-000, Juazeiro do Norte, Cear\'{a}, Brazil}

\author{A. C. A. Ramos}
\affiliation{Centro de Ci\^{e}ncias e Tecnologia, Universidade Federal do Cariri, 57072-270, Juazeiro do Norte, Cear\'{a}, Brazil}

\date{\today}

\begin{abstract}
In this paper we study the corrections emergent from a Ho\v{r}ava-Lifshitz extension of the complex scalar sector to the Bose-Einstein condensation and to the thermodynamics parameters. We initially discussed some features of the model to only then compute the corrections to the Bose-Einstein condensation. The calculations were done by computing the generating functional, from which we extract the thermodynamics parameters. We also obtained the Lifshitz scaling correction for the critical temperature $T_c$ that sets the Bose-Einstein Condensation.
\end{abstract}

\maketitle

\section{Introduction}

Fundamental problems related to the primordial universe or black holes, for example, require a conciliation between quantum mechanics and general relativity, i.e., a quantum description of gravity. String theory \cite{polchinski} is one of the main candidates to provide such quantum description for gravity. However Ho\v{r}ava-Lifhitz (HL) theory of gravity has gained attention in recent years \cite{Horava:2009uw} since it provides a theoretical framework to study quantum gravity in a self contained manner in (3+1) space time dimensions. 

Ho\v{r}ava proposed a new model for quantum gravity \cite{Horava:2009uw} based on an anisotropy between space and time. Such anisotropy breaks, therefore, the Lorentz symmetry and it is very common in condensed matter systems \cite{Lifshitz, Lifshitz2, Sachdev}. The model is built in order to be compatible with the anisotropic scaling,
\begin{equation}
    \vec{x}\rightarrow b\vec{x},\,\,\,\,\,t\rightarrow b^z t,
\end{equation}
where the parameter $z$ which controls the degree of anisotropy is called critical exponent. The anisotropic scaling reflects on the mass dimensions of all fields, operators and coupling constants, so that,
\begin{equation}
    [\vec{x}]=-1,\,\,\,\,\,[t]=-z.
\end{equation}
The great interest in HL model for gravity lies in the fact that the gravity becomes power counting renormalizable when $z=3$, since the dimension of the gravitational coupling constant goes to zero. After Ho\v{r}ava's seminal paper \cite{Horava:2009uw}, several studies were done within the context of gravity \cite{DOdorico:2015pil, Lopes:2015bra, Hartong:2016yrf, Hartong:2015zia}, black holes \cite{Wei:2019zdf, Xu:2018tdh, Xu:2020xsl, AyonBeato:2009nh}, cosmology \cite{Kiritsis:2009sh, Calcagni:2009ar, Oliveira-Neto:2019tvo, Pitelli:2012sj}, holography \cite{Griffin:2012qx, Christensen:2013lma, Mohammadi:2020jbe} and so on.

More recently other HL-like field theories were constructed. Extensions of QED with lifshitz scaling were studied specially in the context of radiative corrections and renormalization \cite{Gomes:2018umk, Gomes:2016ixw}. HL-like extensions for Gross-Neveu \cite{Lima:2016qgr} and four-fermions \cite{Gomes:2020nbz, Mariz:2019kmc, Mariz:2019iij} models were also studied. The finite temperature effects were studied in HL-like extensions of the scalar QED and for Yukawa coupling \cite{Farias:2013rya}.  

A Bose-Einstein condensation (BEC) \cite{Einstein} occurs when a low density gas of bosons is cooled to temperatures near to zero kelvin. Under such conditions a considerable portion of the bosons fall into the lowest quantum state, forming therefore a single collective quantum wave called Bose-Einstein condensate \cite{Huang}. After the first Bose-Einstein condensates produced with rubidium \cite{Anderson} and sodium atoms \cite{Davis}, several studies were carried out in the context quasi-particles \cite{Deng, Kasprzak, Balili, Demokritov}, photons gases \cite{Klaers1} and molecules \cite{Klaers2}.

Bose-Einstein condensates has been intensively studied in the context of high energy physics as well. The critical temperature of Bose-Einstein condensation in Rindler space has been analyzed in \cite{Takeuchi:2015nga}. The effects of Lorentz symmetry violation in the BEC were studied in \cite{Furtado:2020olp, Casana:2011bv, deSales:2011jy, Castellanos:2010zz, Colladay:2006rt}. Also brane-world scenarios \cite{Casadio:2016aum, Kasamatsu:2013lda}, cosmology \cite{Erdem:2019gpk, Atazadeh:2016zls} and black holes \cite{Garay:1999sk, Kuhnel:2015qaa} were analyzed in connection with BEC. 

In this paper we study the corrections emergent from a Ho\v{r}ava-Lifshitz extension of the complex scalar sector to BEC and to the thermodynamics parameters. We initially discussed some features of the model to only then compute the corrections to the BEC. The calculations were done by computing the generating functional, from which we extract the thermodynamics parameters. We also obtained the Lifshitz scaling correction for the critical temperature $T_c$ that sets the Bose-Einstein condensation.

This paper is organized as follows: in the next section we present the model itself and we compute the thermodynamic parameters, such as pressure, energy, specific heat and charge density when $z=2$, which is the simplest non-trivial case. In section III we extend the calculations of the section II for the case of a generic $z$. We also calculate the critical temperature for the BEC. In section IV we present our final remarks. 

\section{Case $z=2$}\label{casez2}
The model we are considering consists of the Ho\v{r}ava-Lifshitz extension of the scalar sector with critical exponent $z=2$. Hence the lagrangian describing the system is

\begin{eqnarray}\label{lagrangian1}
\mathcal{L}&=&\partial_0\phi(\partial_0\phi)^*-\partial_i\partial_j\phi(\partial_i\partial_j\phi)^*-m^4\phi\phi^*.
\end{eqnarray}
In the present model the lagrangian has mass dimension $d=5$ and the mass dimension of the parameters are the following:
\begin{eqnarray}\label{massdimensions}
[\phi]=\frac{1}{2},\,\,[m]=1.
\end{eqnarray}

The lagrangian (\ref{lagrangian1}) possess an obvious $U(1)$ symmetry, so that
\begin{equation}
    \phi\rightarrow \phi'=e^{-i\alpha}\phi,
\end{equation}
with $\alpha\in\mathbb{R}$. The Noether's theorem states that for any given continuous symmetry there is a conserved quantity in connection. In order to find such conserved quantity let us consider $\alpha=\alpha(x)$, i.e., a function of spacetime position, so that the Euler-Lagrange equation gives us the equation of motion for the ``field'' $\alpha(x)$. Since the contribution $\partial\mathcal{L}/\partial\alpha=0$, we can find the following conserved current,
\begin{equation}
    j_{k}(x)=i\left(\partial_m\phi\partial_m\partial_k\phi^*-\partial_m\partial_k\phi\partial_m\phi^*+\phi^*\partial_k\partial_m^2\phi-\phi\partial_k\partial_m^2\phi^*\right),
\end{equation}
and charge density
\begin{eqnarray}
    \nonumber Q&=&\int d^3x j^0\\
    &=&\int d^3x\left[i\left(\phi^*\frac{\partial\phi}{\partial t}-\phi\frac{\partial\phi^*}{\partial t}\right)\right].
\end{eqnarray}

Making use of the equations of motion for $\phi$ and $\phi^*$, given by
\begin{eqnarray}
-\partial_0^2\phi-\partial_m^2\partial_n^2\phi-m^4\phi&=&0\\
-\partial_0^2\phi^*-\partial_m^2\partial_n^2\phi^*-m^4\phi^*&=&0,
\end{eqnarray}
a straightforward calculation shows us the conservation of the four current, i.e., $\partial_{\mu}j^{\mu}(x)=0$. Splitting the fieds $\phi$ and $\phi^*$ into two real components $\phi_1$ and $\phi_2$ as
\begin{eqnarray}
    \phi&=&\frac{1}{\sqrt{2}}(\phi_1+i\phi_2)\\
    \phi^*&=&\frac{1}{\sqrt{2}}(\phi_1-i\phi_2),
\end{eqnarray}
we can rewrite the Lagrangian in terms of $\phi_a$ with $a=1,2$ as follows
\begin{eqnarray}
    \mathcal{L}&=&\frac{1}{2}\partial_{0}\phi_a\partial_{0}\phi_a-\frac{1}{2}\partial_{i}\partial_{j}\phi_a\partial_{i}\partial_{j}\phi_a-\frac{m^4}{2}\phi_a\phi_a.
\end{eqnarray}
The canonically conjugated momenta are:
\begin{eqnarray}\label{momenta}
    \pi_a=\partial_0\phi_a
\end{eqnarray}
then the Hamiltonian becomes
\begin{eqnarray}
    \mathcal{H}=\frac{1}{2}\left(\pi_a\pi_a+\partial_{i}\partial_{j}\phi_a\partial_{i}\partial_{j}\phi_a+m^4\phi_a\phi_a\right).
\end{eqnarray}
The charge density can also be expressed in terms of $\phi_a$, 
\begin{equation}\label{chargedensity1}
    Q=\int d^3x\epsilon_{ab}\pi_a\phi_b.
\end{equation}
Letting $\mathcal{H}(\phi,\pi)\rightarrow\mathcal{H}(\phi,\pi)-\mu\mathcal{N}(\phi,\pi)$, where $\mathcal{N}(\phi,\pi)$ is the conserved charge density, identified as $Q$, and $\mu$ is the chemical potential, the partition function becomes:
\begin{eqnarray}\label{Z1}
    \nonumber Z&=&\int D\pi_a\int_{periodic}D\phi_a\exp\left\{\int_0^{\beta}d\tau\int d^3x\times\right.\\
    &&\times\left.\left[i\pi_a\frac{\partial\phi_a}{\partial\tau}-\mathcal{H}(\phi_a,\pi_a)+\mu\epsilon_{ab}\pi_a\phi_b\right]\right\}.
\end{eqnarray}
Note that the $[\mu]=2$. The term ``periodic'' states for the fact that the integration over the field is constrained in such way that $\phi(\vec{x},0)=\phi(\vec{x},\beta)$ with $\beta=1/T$. The partition function can be written as
\begin{eqnarray}\label{Z2}
    \nonumber Z&=&\int D\pi_a\int_{periodic}D\phi_a\exp\left\{\int_0^{\beta}d\tau\int d^3x\times\right.\\
    \nonumber&&\times\left.\left[-\frac{1}{2}\pi_a^2+\left(i\frac{\partial\phi_a}{\partial\tau}-\mu\epsilon_{ab}\phi_b\right)\pi_a\right.\right.\\
    &&\left.\left.-\frac{1}{2}\partial_{i}\partial_{j}\phi_a\partial_{i}\partial_{j}\phi_a-\frac{1}{2}m^4\phi_a^2\right]\right\}.
\end{eqnarray}

The integration over the momenta can be directly done. Then we obtain,
\begin{eqnarray}\label{Z3}
    \nonumber Z&=&(N')^2\int_{periodic}D\phi_i\exp\left\{\int_0^{\beta}d\tau\int d^3x\times\right.\\
    \nonumber&&\times\left.\left[\frac{1}{2}\left(i\frac{\partial\phi_b}{\partial\tau}-\mu\epsilon_{ab}\phi_a\right)^2-\frac{1}{2}\partial_{i}\partial_{j}\phi_a\partial_{i}\partial_{j}\phi_a-\frac{1}{2}m^4\phi_a^2\right]\right\}.\\
\end{eqnarray}
The factor $N'$ is an irrelevant normalization constant, since multiplication of $Z$ by any constant does not change the thermodynamics. The components of $\phi$ can be Fourier-expanded as,
\begin{eqnarray}\label{Fourier}
    \phi_1&=&\sqrt{2}\zeta\cos\theta+\sqrt{\frac{\beta}{V}}\sum_{n}\sum_{\vec{p}}e^{i(\vec{p}\cdot\vec{x}+\omega_n\tau)}\phi_{1;n}(\vec{p})\\
    \phi_2&=&\sqrt{2}\zeta\sin\theta+\sqrt{\frac{\beta}{V}}\sum_{n}\sum_{\vec{p}}e^{i(\vec{p}\cdot\vec{x}+\omega_n\tau)}\phi_{2;n}(\vec{p}),
\end{eqnarray}
where $\omega_n=2\pi n T$, owing to the constraint of periodicity that $\phi(\vec{x},\beta)=\phi(\vec{x},0)$ for all $\vec{x}$. Here $\zeta$ and $\theta$ are independent of spacetime position and determine the full infrared behaviour of the field; that is, $\phi_{1;0}(\vec{p}=\vec{0})=\phi_{1;0}(\vec{p}=\vec{0})=0$. This allows for the possibility of condensation of the bosons into the zero-momentum state. Substituting (\ref{Fourier}) into (\ref{Z2}) the partition function becomes
\begin{equation}
    Z=(N')^2\prod_n\prod_p\int D\phi_{1;n}(\vec{p})D\phi_{2;n}(\vec{p})e^{S},
\end{equation}
where $S$ is given by

\begin{eqnarray}
    \nonumber S&=&\beta V(\mu^2-m^4)\zeta^2\\
    &&-\frac{1}{2}\sum_n\sum_p\left(\phi_{1;-n}(-\vec{p}),\phi_{2;-n}(-\vec{p})\right)D\left(\begin{array}{c}
\phi_{1;n}(\vec{p})   \\
\phi_{1;n}(\vec{p})
    \end{array}\right),
\end{eqnarray}
being $D$

\begin{eqnarray}
    D=\beta^2\left(\begin{array}{cc}
    \omega_n^2+\omega^2-\mu^2 & -2\mu\omega_n \\
    2\mu\omega_n & \omega_n^2+\omega^2-\mu^2
    \end{array}\right),
\end{eqnarray}
with $\omega=\sqrt{\vec{p}^4+m^4}$. Carrying out the integrations over $\phi_{1;n}$ and $\phi_{2;n}$, we have,
\begin{equation}
    \ln Z=\beta V(\mu^2-m^4)\zeta^2+\ln (\det D)^{-1/2},
\end{equation}
so that we can rewritte in the following form

\begin{eqnarray}\label{generatingfunctional}
    \nonumber\ln Z&=&\beta V(\mu^2-m^4)\zeta^2\\
    \nonumber&&-V\int\frac{d^3p}{(2\pi)^3}\left\{\beta\omega+\ln\left[1-e^{-\beta(\omega-\mu)}\right]+\ln\left[1-e^{-\beta(\omega+\mu)}\right]\right\}\\
\end{eqnarray}

It is important to highlight here that the above expression for $\ln Z$ was obtained under the consideration of the following convergence condition
\begin{equation}\label{convergence}
    \left|\mu\right|\leq m^2.
\end{equation}
A modification of the usual convergence condition ($|\mu|\leq m$) already known in the literature, first stated in the work of Haber \cite{Haber:1981fg}. 
The usual relation 
\begin{equation}
    \frac{PV}{T}=\ln Z,
\end{equation}
gives us the equation of state for the system. Hence the pressure is given by
\begin{eqnarray}\label{pressurecomplete}
    \nonumber P&=&(\mu^2-m^4)\zeta^2\\
    \nonumber&&-\frac{1}{\beta}\int\frac{d^3p}{(2\pi)^3}\left\{\beta\omega+\ln\left[1-e^{-\beta(\omega-\mu)}\right]+\ln\left[1-e^{-\beta(\omega+\mu)}\right]\right\}.\\
\end{eqnarray}
The internal energy can be written as
\begin{eqnarray}\label{energy}
    \nonumber E&=&-\frac{\partial}{\partial \beta}\ln Z\\
   \nonumber &=&-V(\mu^2-m^4)\zeta^2\\
    &&-V\int\frac{d^3p}{(2\pi)^3}\left\{-\omega+\frac{-\mu-\omega}{e^{\beta(\mu+\omega)}-1}-\frac{\mu-\omega}{e^{\beta(\omega-\mu)}-1}\right\}.
\end{eqnarray}
The specific heat at constant volume is expressed by

\begin{eqnarray}\label{heat}
    \nonumber C_v&=&\frac{\partial E}{\partial T}\\
    \nonumber&=&-V\beta^2\int\frac{d^3p}{(2\pi)^3}\left\{\frac{(\mu-\omega)(\omega-\mu)e^{-\beta(\omega-\mu)}}{1-e^{-\beta(\omega-\mu)}}\right.\\
    \nonumber&&+\left.\frac{(\mu-\omega)(\omega-\mu)e^{-2\beta(\omega-\mu)}}{\left(1-e^{-\beta(\omega-\mu)}\right)^2}\right.\\
    \nonumber&&+\frac{(-\mu-\omega)(\omega+\mu)e^{-\beta(\omega+\mu)}}{1-e^{-\beta(\omega+\mu)}}\\
    &&+\left.\frac{(-\mu-\omega)(\omega+\mu)e^{-2\beta(\omega+\mu)}}{\left(1-e^{-\beta(\omega+\mu)}\right)^2}\right\}
\end{eqnarray}
The charge density is written as

\begin{eqnarray}\label{charge}
    \nonumber\rho&=&\frac{1}{\beta V}\frac{\partial \ln Z}{\partial \mu}\\
     &=& 2\mu\zeta^2+\int\frac{d^3p}{(2\pi)^3}\left[\frac{e^{-\beta  (\omega -\mu )}}{1-e^{-\beta  (\omega -\mu )}}-\frac{e^{-\beta  (\mu +\omega )}}{1-e^{-\beta  (\mu +\omega )}}\right]
\end{eqnarray}
Using spherical coordinates, the integration measure goes to

\begin{equation}
    \int\frac{d^3p}{(2\pi)^3}\longrightarrow\frac{1}{2\pi^2}\int dp p^2.
\end{equation}
In a non-condensate fase, by setting $\zeta=0$, the expressions for the above thermodynamic quantities can be simplified. So that we obtain for the pressure

\begin{eqnarray}\label{pressurecomplete1}
    \nonumber P=-\frac{1}{\beta}\frac{1}{2\pi^2}\int dp p^2\left\{\beta\omega+\ln\left[1-e^{-\beta(\omega-\mu)}\right]+\ln\left[1-e^{-\beta(\omega+\mu)}\right]\right\},\\
\end{eqnarray}
for the energy
\begin{eqnarray}\label{energy2}
     E=-V\frac{1}{2\pi^2}\int dp p^2\left\{-\omega+\frac{-\mu-\omega}{e^{\beta(\mu+\omega)}-1}-\frac{\mu-\omega}{e^{\beta(\omega-\mu)}-1}\right\},
\end{eqnarray}
for the specific heat at constant volume
\begin{eqnarray}\label{heat2}
    \nonumber C_v&=&-V\beta^2\frac{1}{2\pi^2}\int dp p^2\left\{\frac{(\mu-\omega)(\omega-\mu)e^{-\beta(\omega-\mu)}}{1-e^{-\beta(\omega-\mu)}}\right.\\
    \nonumber&&+\left.\frac{(\mu-\omega)(\omega-\mu)e^{-2\beta(\omega-\mu)}}{\left(1-e^{-\beta(\omega-\mu)}\right)^2}\right.\\
    \nonumber&&+\frac{(-\mu-\omega)(\omega+\mu)e^{-\beta(\omega+\mu)}}{1-e^{-\beta(\omega+\mu)}}\\
    &&+\left.\frac{(-\mu-\omega)(\omega+\mu)e^{-2\beta(\omega+\mu)}}{\left(1-e^{-\beta(\omega+\mu)}\right)^2}\right\},
\end{eqnarray}
and finally for the charge density
\begin{eqnarray}\label{charge2}
    \rho&=&\frac{1}{2\pi^2}\int dp p^2\left[\frac{e^{-\beta  (\omega -\mu )}}{1-e^{-\beta  (\omega -\mu )}}-\frac{e^{-\beta  (\mu +\omega )}}{1-e^{-\beta  (\mu +\omega )}}\right].
\end{eqnarray}

At this point it is important to discuss the influence of the critical exponent $z=2$ in the critical temperature $T_c$ that sets the Bose-Einstein condensation. For the occurrence of the Bose-Einstein condensation we must have $\mu=\pm m^2$. Applying such condition in (\ref{charge2}) we obtain
\begin{equation}
    T_c=\sqrt{\frac{3\rho}{m^2}},
\end{equation}
which is the correction for the critical temperature first presented in \cite{Haber:1981fg}.

\section{Arbitrary $z$ case}

For a generic critical exponent $z$ we have the following lagrangian.

\begin{eqnarray}\label{lagrangianz}
\mathcal{L}=\partial_0\phi(\partial_0\phi)^*-\partial_{i1}\partial_{i2}\cdots\partial_{iz}\phi(\partial_{i1}\partial_{i1}\cdots\partial_{iz}\phi)^*-m^{2z}\phi\phi^*.
\end{eqnarray}
The mass dimension of the lagrangian is $d+z$ being $d$ the spatial dimension which in our work is considered as $d=3$. In this case the spatial derivative has dimension $[\partial_i]=1$ while the temporal derivative has dimension $[\partial_0]=z$. The mass dimension of the scalar field and the mass parameter are the same as presented in (\ref{massdimensions}).

\begin{figure*}[ht!]
    \centering
    \includegraphics[scale=0.6]{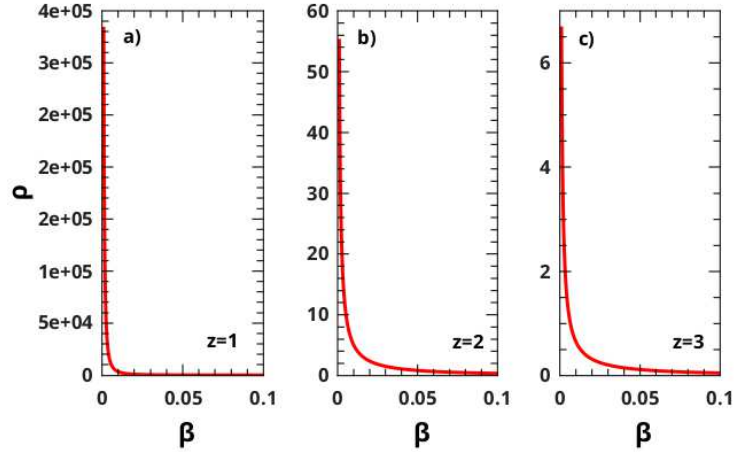}
    \caption{Charge density for $z=1$, $z=2$ and $z=3$. In this plot we have considered $\mu=1$ and $m=2$.}
    \label{fig1}
\end{figure*}

The same procedure employed in section (\ref{casez2}) can be applied here so that we can found the Hamiltonian density as:
\begin{eqnarray}
    \mathcal{H}=\frac{1}{2}\left(\pi_a\pi_a+\partial_{i1}\partial_{i2}\cdots\partial_{iz}\phi_a\partial_{i1}\partial_{i2}\cdots\partial_{iz}\phi_a+m^{2z}\phi_a\phi_a\right).
\end{eqnarray}
The charge density and the momenta are kept unchanged, i.e., they are given by (\ref{momenta}) and (\ref{chargedensity1}). Thence the generating functional can be written as follows

\begin{eqnarray}\label{Zz}
    \nonumber Z&=&\int D\pi_a\int_{periodic}D\phi_a\exp\left\{\int_0^{\beta}d\tau\int d^3x\times\right.\\
    \nonumber&&\times\left.\left[-\frac{1}{2}\pi_a^2+\left(i\frac{\partial\phi_a}{\partial\tau}-\mu\epsilon_{ab}\phi_b\right)\pi_a\right.\right.\\
    &&\left.\left.-\frac{1}{2}\partial_{i1}\partial_{i2}\cdots\partial_{iz}\phi_a\partial_{i1}\partial_{i2}\cdots\partial_{iz}\phi_a-\frac{1}{2}m^{2z}\phi_a^2\right]\right\}.
\end{eqnarray}
The chemical potential for the generic $z$ case has mass dimension $[\mu]=z$. The final form of the partition function after the integrations and the sum over the Matsubara frequencies is same as presented in (\ref{generatingfunctional}), where the only differences are the expression for the dispersion relation, which is given by $\omega=\sqrt{\vec{p}^{2z}+m^{2z}}$ and the term related to the condensed fase $\beta V(\mu^2-m^{2z})\zeta^2$. Consequently, the expressions for the pressure, energy, specific heat and charge density in the non-condensate fase are also the same as those presented in equations (\ref{pressurecomplete1}), (\ref{energy2}), (\ref{heat2}) and (\ref{charge2}) respectively.

The convergence condition is also modified by the critical exponent $z$. For a generic $z$ we have the following convergence condition
\begin{equation}\label{convergenceZ}
    |\mu|\leq m^{z},
\end{equation}
which restores the usual convergence criteria first stated by Haber \cite{Haber:1981fg} when $z=1$. From this corrected convergence criteria we can obtain, with the same procedure used in the previous section, the critical temperature for an arbitrary critical exponent $z$. Therefore from (\ref{convergenceZ}) we obtain
\begin{equation}\label{Tcz}
    T_c=\sqrt{\frac{3\rho}{m^z}}.
\end{equation}
As we can see from (\ref{Tcz}), as we increase the critical exponent the critical temperature for the Bose-Einstein condensation decreases.

\begin{figure*}[ht!]
    \centering
    \includegraphics[scale=0.6]{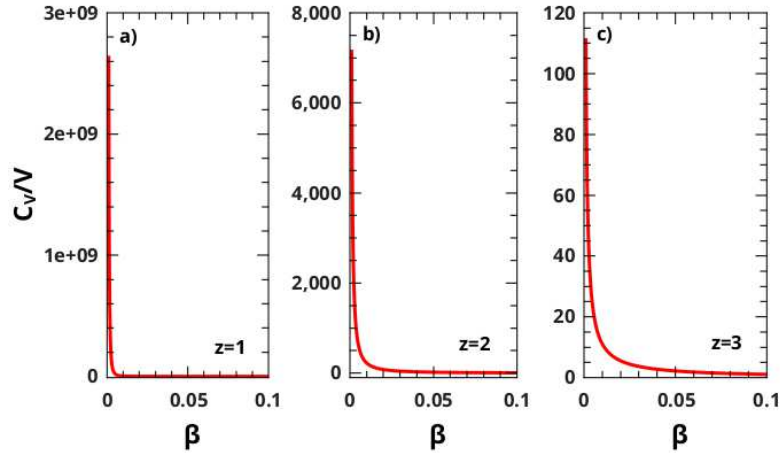}
    \caption{Specific heat for volume unit for $z=1$, $z=2$ and $z=3$. In this plot we have considered $\mu=1$ and $m=2$.}
    \label{fig2}
\end{figure*}

We plot the qualitative behaviour of the charge density (fig. \ref{fig1}) and specific heat at constant volume (fig. \ref{fig2}) for three values of the critical exponent $z$, namely, $z=1$, $z=2$ and $z=3$. We have considered for this plot the chemical potential $\mu=1$ and $m=2$. As we can see for both charge density and specific heat, the increasing of $z$ turns the divergence around $\beta=0$ softer. Also when $\beta\rightarrow \infty$ ($T\rightarrow 0$) the effect of $z$ becomes negligible. 

\section{Final Remarks}

In this paper we study the corrections emergent from a Ho\v{r}ava-Lifshitz extension of the complex scalar sector to the Bose-Einstein condensation and to the thermodynamics parameters. We initially discussed some features of the model to only then compute the corrections to the Bose-Einstein condensation. The calculations were done by computing the generating functional, from which we extract the thermodynamic parameters. We also obtained the Lifshitz scaling correction for the critical temperature $T_c$ that sets the Bose-Einstein Condensation.

We considered initially the first non-trivial case of critical exponent, i.e., $z=2$. For $z=2$ case we calculate the analytical expressions for pressure, energy, specific heat and charge density. Besides we also obtained a correction for the convergence criteria for the generating functional from which we compute the Bose-Einstein condensation critical temperature.

We also addressed the generic $z$ case, from which we obtain the pressure, energy specific heat and charge density, as well as the Bose-Einstein condensation critical temperature. We could see that increasing the value of $z$ the Bose-Einstein critical temperature decreases. A natural continuation of this work consists in the consideration of self interacting scalar fields.




\begin{thebibliography}{99}

\bibitem{polchinski}
 J. Polchinski, String Theory Vol. I Cambridge University Press (1999)

\bibitem{Horava:2009uw}
P.~Horava,
Phys. Rev. D \textbf{79} (2009), 084008

\bibitem{Lifshitz}
E.M. Lifshitz, Zh.Eksp.Teor.Fiz. 11 (1941) 255

\bibitem{Lifshitz2}
E.M. Lifshitz, Zh.Eksp.Teor.Fiz. 11 (1941) 269

\bibitem{Sachdev}
S. Sachdev, Quantum Phase Transitions, Cambridge University Press (1999)

\bibitem{DOdorico:2015pil}
G.~D'Odorico, J.~W.~Goossens and F.~Saueressig,
JHEP \textbf{10} (2015), 126

\bibitem{Lopes:2015bra}
D.~V.~Lopes, A.~Mamiya and A.~Pinzul,
Class. Quant. Grav. \textbf{33} (2016) no.4, 045008

\bibitem{Hartong:2016yrf}
J.~Hartong, Y.~Lei and N.~A.~Obers,
Phys. Rev. D \textbf{94} (2016) no.6, 065027

\bibitem{Hartong:2015zia}
J.~Hartong and N.~A.~Obers,
JHEP \textbf{07} (2015), 155

\bibitem{Wei:2019zdf}
S.~W.~Wei, J.~Yang and Y.~X.~Liu,
Phys. Rev. D \textbf{99} (2019) no.10, 104016

\bibitem{Xu:2018tdh}
J.~Xu and J.~Jing,
Annals Phys. \textbf{389} (2018), 136-147

\bibitem{Xu:2020xsl}
H.~Xu and Y.~C.~Ong,
Eur. Phys. J. C \textbf{80} (2020) no.7, 679

\bibitem{AyonBeato:2009nh}
E.~Ayon-Beato, A.~Garbarz, G.~Giribet and M.~Hassaine,
Phys. Rev. D \textbf{80} (2009), 104029

\bibitem{Kiritsis:2009sh}
E.~Kiritsis and G.~Kofinas,
Nucl. Phys. B \textbf{821} (2009), 467-480
  
\bibitem{Calcagni:2009ar}
G.~Calcagni,
JHEP \textbf{09} (2009), 112

\bibitem{Oliveira-Neto:2019tvo}
G.~Oliveira-Neto, L.~G.~Martins, G.~A.~Monerat and E.~V.~Corr\^ea Silva,
Int. J. Mod. Phys. D \textbf{28} (2019) no.10, 1950130

\bibitem{Pitelli:2012sj}
J.~P.~M.~Pitelli and A.~Saa,
Phys. Rev. D \textbf{86} (2012), 063506

\bibitem{Griffin:2012qx}
T.~Griffin, P.~Ho\v{r}ava and C.~M.~Melby-Thompson,
Phys. Rev. Lett. \textbf{110} (2013) no.8, 081602

\bibitem{Christensen:2013lma}
M.~H.~Christensen, J.~Hartong, N.~A.~Obers and B.~Rollier,
Phys. Rev. D \textbf{89} (2014), 061901

\bibitem{Mohammadi:2020jbe}
M.~Mohammadi and A.~Sheykhi,
Eur. Phys. J. C \textbf{80} (2020) no.10, 928

\bibitem{Gomes:2018umk}
M.~Gomes, F.~Marques, T.~Mariz, J.~R.~Nascimento, A.~Y.~Petrov and A.~J.~da Silva,
Phys. Rev. D \textbf{98} (2018) no.10, 105016

\bibitem{Gomes:2016ixw}
M.~Gomes, T.~Mariz, J.~R.~Nascimento, A.~Y.~Petrov and A.~J.~da Silva,
Phys. Lett. B \textbf{764} (2017), 277-281

\bibitem{Lima:2016qgr}
A.~M.~Lima, T.~Mariz, R.~Martinez, J.~R.~Nascimento, A.~Y.~Petrov and R.~F.~Ribeiro,
Phys. Rev. D \textbf{95} (2017) no.6, 065031

\bibitem{Gomes:2020nbz}
M.~Gomes, T.~Mariz, J.~R.~Nascimento, A.~Y.~Petrov and A.~J.~da Silva,
Eur. Phys. J. C \textbf{80} (2020) no.6, 518

\bibitem{Mariz:2019kmc}
T.~Mariz, J.~R.~Nascimento and A.~Y.~Petrov,
Phys. Rev. D \textbf{101} (2020) no.10, 105008

\bibitem{Mariz:2019iij}
T.~Mariz, R.~Moreira and A.~Y.~Petrov,
Eur. Phys. J. C \textbf{79} (2019) no.7, 550
[erratum: Eur. Phys. J. C \textbf{79} (2019) no.9, 729]

\bibitem{Farias:2013rya}
C.~F.~Farias, M.~Gomes, J.~R.~Nascimento, A.~Y.~Petrov and A.~J.~da Silva,
Phys. Rev. D \textbf{89} (2014) no.2, 025014
doi:10.1103/PhysRevD.89.025014
[arXiv:1311.6313 [hep-th]].

\bibitem{Einstein}
A. Einstein, Zweite Abhandlung. Sitz. ber. Preuss. Akad. Wiss. {\bf 1} (1925), 3–14

\bibitem{Huang}
Huang, K. Statistical Mechanics 2nd edn 293–294 (Wiley, 1987).

\bibitem{Anderson}
M. H. Anderson, J. R. Ensher, M. R. Matthews, C. E Wieman, E. A. Cornell,
Science {\bf 269} (1995), 198–201.

\bibitem{Davis}
K. B. Davis, et al., 
Phys. Rev. Lett. {\bf 75}, (1995), 3969–3973.

\bibitem{Deng}
H. Deng, G. Weihs, C. Santori, J. Bloch, Y. Yamamoto, 
Science {\bf 298} (2002), 199–202.

\bibitem{Kasprzak}
J. Kasprzak, et al.  
Nature {\bf 443} (2006), 409–414.

\bibitem{Balili}
R. Balili, V. Hartwell, D. Snoke, L. Pfeiffer, K. West, 
Science {\bf 316} (2007), 1007–1010.

\bibitem{Demokritov}
S. O. Demokritov, et al.  
Nature {\bf 443}, (2006) 430–433.

\bibitem{Klaers1}
J. Klaers, F. Vewinger, M. Weitz.  
Nature Phys. {\bf 6} (2010), 512–515.

\bibitem{Klaers2}
J. Klaers, J. Schmitt, F. Vewinger, et al.  
Nature {\bf 468} (2010), 545–548.

\bibitem{Takeuchi:2015nga}
S.~Takeuchi,
Phys. Lett. B \textbf{750} (2015), 209-217
[arXiv:1501.07471 [hep-th]].

\bibitem{Furtado:2020olp}
J.~Furtado, A.~C.~A.~Ramos and J.~F.~Assun\c{c}\~ao,
EPL \textbf{132} (2020) no.3, 31001
[arXiv:2009.07034 [hep-th]].

\bibitem{Casana:2011bv}
R.~Casana and K.~A.~T.~da Silva,
Mod. Phys. Lett. A \textbf{30} (2015) no.07, 1550037
[arXiv:1106.5534 [hep-th]].

\bibitem{deSales:2011jy}
J.~A.~de Sales, T.~Costa-Soares and V.~J.~V.~Otoya,
Physica A \textbf{391} (2012), 5422-5432
[arXiv:1106.4604 [hep-th]].

\bibitem{Castellanos:2010zz}
E.~Castellanos and A.~Camacho,
Mod. Phys. Lett. A \textbf{25} (2010), 459-469

\bibitem{Colladay:2006rt}
D.~Colladay and P.~McDonald,
Phys. Rev. D \textbf{73} (2006), 105006
[arXiv:hep-ph/0602071 [hep-ph]].

\bibitem{Casadio:2016aum}
R.~Casadio and R.~da Rocha,
Phys. Lett. B \textbf{763} (2016), 434-438
[arXiv:1610.01572 [hep-th]].

\bibitem{Kasamatsu:2013lda}
K.~Kasamatsu, H.~Takeuchi and M.~Nitta,
J. Phys. Condens. Matter \textbf{25} (2013), 404213
[arXiv:1303.4469 [cond-mat.quant-gas]].

\bibitem{Erdem:2019gpk}
R.~Erdem and K.~G\"ultekin,
JCAP \textbf{10} (2019), 061
[arXiv:1908.08784 [gr-qc]].

\bibitem{Atazadeh:2016zls}
K.~Atazadeh, F.~Darabi and M.~Mousavi,
Eur. Phys. J. C \textbf{76} (2016) no.6, 327
[arXiv:1601.07516 [gr-qc]].

\bibitem{Garay:1999sk}
L.~J.~Garay, J.~R.~Anglin, J.~I.~Cirac and P.~Zoller,
Phys. Rev. Lett. \textbf{85} (2000), 4643-4647
[arXiv:gr-qc/0002015 [gr-qc]].

\bibitem{Kuhnel:2015qaa}
F.~K\"uhnel and M.~Sandstad,
Phys. Rev. D \textbf{92} (2015) no.12, 124028
[arXiv:1506.08823 [gr-qc]].

\bibitem{Haber:1981fg}
  H.~E.~Haber and H.~A.~Weldon,
  Phys.\ Rev.\ Lett.\  {\bf 46} (1981) 1497.
  

     

\end{thebibliography}
\end{document}